\documentclass[article,pre,10pt,twocolumn,longbibliography]
{revtex4-1}
\usepackage{microtype}
\usepackage{graphicx}
\usepackage[colorlinks,linkcolor=blue,urlcolor=blue,citecolor=blue]{hyperref}

\usepackage{caption}
\usepackage{subcaption}
\usepackage{mathptmx}
\usepackage{times}
\usepackage{epsfig,amsopn}
\usepackage{graphicx}
\usepackage{bm,amsmath,amssymb}
\usepackage{natbib}
\usepackage{braket}
\usepackage{float}
\usepackage{wasysym}
\usepackage{enumerate}
\usepackage{hyperref}
\usepackage{comment}
\allowdisplaybreaks[1]
\usepackage[dvipsnames]{xcolor}

\def\be{\begin{equation}}
\def\ee{\end{equation}}
\def\bse{\begin{subequations}}
\def\ese{\end{subequations}}
\def\bcs{\begin{cases}}
\def\ecs{\end{cases}}
\def\bea{\begin{eqnarray}}
\def\eea{\end{eqnarray}}

\newcommand{\Aref}[1]{Appendix}%

\newcommand{\opunit}{\textrm{1}\kern-0.22em\textrm{l}}

\begin{document}

\title{Discontinuity in the distribution of field increments between avalanches in non-abelian  random field Blume-Emery-Griffiths model with no passing violation}

\author{{\normalsize{}Aldrin B E$^{1, 2}$}
{\normalsize{}}}

\author{{\normalsize{}Alberto Rosso$^{3}$}
{\normalsize{}}}

\author{{\normalsize{}Sumedha$^{1, 2}$}
{\normalsize{}}}
\email{sumedha@niser.ac.in}

\affiliation{\noindent $^{1}$School of Physical Sciences, National Institute of Science Education and Research, Bhubaneswar, P.O. Jatni, 752050, India}

\affiliation{\noindent $^{2}$Homi Bhabha National Institute, Training School Complex, Anushakti Nagar 400094, India}

\affiliation{\noindent $^{3}$Universit\'e Paris-Saclay, CNRS, LPTMS, 91405, Orsay, France}

\begin{abstract}
We study the zero-temperature quasi-statically driven dynamics of the random field Blume--Emery--Griffiths model (RFBEGM) as a minimal framework to investigate the consequences of violating the no-passing property in driven disordered systems. 
While the random field Ising model obeys no-passing and displays abelian relaxation dynamics, we show that this property is generically violated in the RFBEGM. 
By systematically exploring the full parameter space of the fully connected model, we identify the regimes in which no-passing is broken and demonstrate that, when this violation is combined with frustration induced by a repulsive biquadratic coupling, it leaves a clear dynamical signature. 
Specifically, the distribution of the minimal field increment required to trigger successive avalanches develops a discontinuity that is absent both in no-passing dynamics and in unfrustrated no-passing-violating regimes. 
We provide analytical arguments that locate the onset of this discontinuity, in excellent agreement with numerical simulations. 
Our results establish this discontinuity as a robust diagnostic of frustration-induced blocking in non-abelian avalanche dynamics within a mean-field setting, without making claims about new universality classes.
\end{abstract}

\date{\today}
\maketitle

\section{Introduction}
\label{Sec1}

The study of driven disordered systems is central to understanding a wide variety of physical phenomena, ranging from crackling noise in magnetic materials~\cite{crackling_noise}, martensitic transformations~\cite{ava_martensites}, and rigidity transitions in granular matter~\cite{rigidity}, to earthquake dynamics along faults~\cite{earthquakes}. 
A hallmark of these systems is that their response to slow external driving is intermittent, proceeding through sudden bursts of activity known as avalanches, reflecting the presence of many metastable states in a complex energy landscape.

A paradigmatic framework to study avalanche dynamics is provided by the athermal random field Ising model (RFIM) driven quasi-statically by an external magnetic field~\cite{sethna,tadic}. 
An important feature of the RFIM dynamics is its abelian character: the final metastable configuration reached after an avalanche does not depend on the order in which unstable spins are updated. 
This property is closely related to the no-passing property (NPP), which ensures that a partial ordering between configurations is preserved under monotonic driving~\cite{middleton,sethna1}.
As a consequence, the dynamics is strongly constrained and reproducible.

However, not all driven systems satisfy these constraints. 
Prominent examples include elasto-plastic models of amorphous solids, where non-monotonic and long-range interactions lead to non-abelian relaxation dynamics~\cite{rossi,lin1,jagla,parley,berthier}. 
This naturally raises a general question: how does the violation of the no-passing property manifest itself in the dynamics of driven disordered systems, and are there robust dynamical signatures that distinguish no-passing and no-passing-violating regimes?

In this work, we address this question using the random field Blume--Emery--Griffiths model (RFBEGM), a three-state extension of the RFIM that includes both bilinear and biquadratic interactions as well as a crystal field term. 
While the RFBEGM has been previously used to model systems such as martensites, ferroelastics, and metamagnets~\cite{beg,aharony,metamag,lawrie,ortin1,ortin2,perez-reche,vasseur}, here we employ it as a minimal and analytically tractable framework that allows one to continuously interpolate between regimes where the no-passing property holds and regimes where it is violated.
Crucially, this interpolation can be achieved while independently controlling the presence or absence of frustration. 

By systematically exploring the parameter space of the fully connected RFBEGM, we show that the violation of the no-passing property alone is not sufficient to qualitatively alter avalanche dynamics. 
Instead, when no-passing violation (NPV) is combined with frustration induced by a repulsive biquadratic coupling, the dynamics develops a distinct feature: a discontinuity in the distribution of the minimal field increment $\delta H_{\min}$ required to trigger successive avalanches.
We demonstrate analytically that the location of this discontinuity can be predicted exactly in the mean-field limit and confirm this prediction numerically.
We interpret this discontinuity as a robust dynamical diagnostic of frustration-induced blocking in non-abelian avalanche dynamics, without making claims about new universality classes.

\begin{figure*}[htbp]
\centering
\includegraphics[width=0.8\hsize]{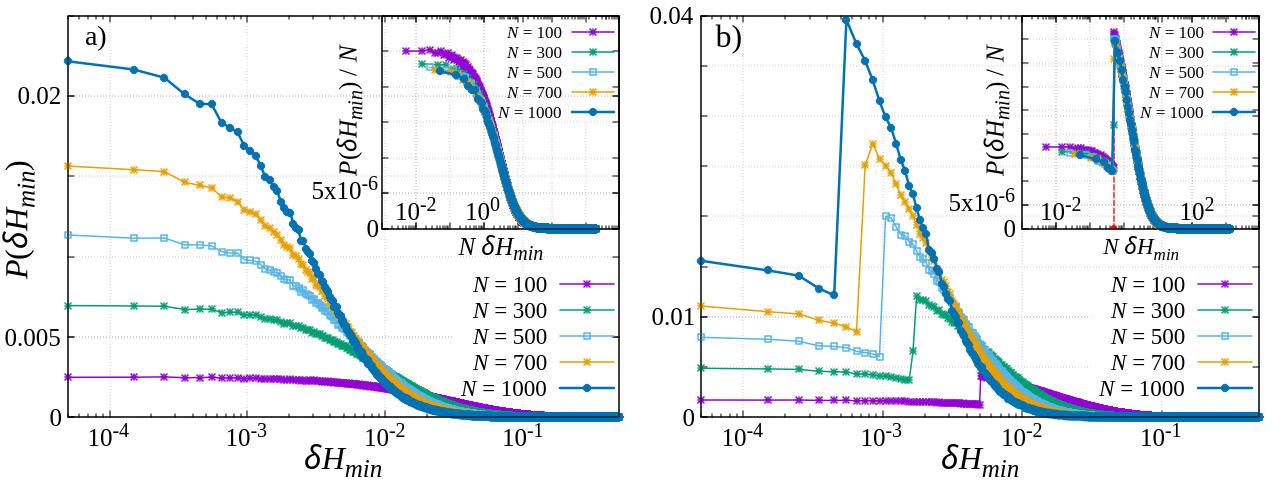}
\caption{
Distribution of the minimal field increment $\delta H_{\min}$ required to destabilize the system from a metastable state. 
(a) No-passing (NPP) regime with $K = -1.0$, $\Delta = -1.5$, where the distribution is continuous and nearly flat for small $\delta H_{\min}$.
(b) No-passing-violating (NPV) regime with repulsive biquadratic coupling ($K = -1.5$, $\Delta = -0.5$), where the distribution develops a clear discontinuity.
Insets show scaling collapses of the form $P(\delta H_{\min}) = N P(N \delta H_{\min})$. 
For the frustrated NPV case, the jump occurs at $N \delta H_{\min} = |K| - 1 = 0.5$, highlighting a qualitative dynamical difference between NPP and frustrated NPV regimes.
}
\label{fig1}
\end{figure*}

This qualitative difference in $P(\delta H_{\min})$ (Fig.~\ref{fig1}) provides a clear dynamical signature of the combined effect of frustration and no-passing violation.
In particular, while no-passing violation alone does not lead to qualitative changes in the field-increment statistics, frustration stabilizes metastable configurations over a finite range of field values, resulting in the observed discontinuity.

The remainder of the paper is organized as follows. 
In Sec.~\ref{Sec2} we define the model and the zero-temperature dynamics. 
In Sec.~\ref{Sec3} we present the dynamical phase diagram in the $K$--$\Delta$ plane and numerical evidence for the discontinuity in the field-increment distribution. 
In Sec.~\ref{Sec4} we provide an analytical argument for the location of the discontinuity. 
In Sec.~\ref{Sec5} we discuss avalanche size statistics, and in Sec.~\ref{Sec6} we conclude with a discussion and outlook.

\section{Model}
\label{Sec2}
The RFBEGM allows us to interpolate continuously between regimes where NPP holds and where it is violated, by tuning two parameters $(K, \Delta)$. We focus on the fully-connected version of the model, defined by the Hamiltonian

\begin{eqnarray}
\mathcal{H}[s] &=& -\frac{J}{2N}\left( \sum_{i}s_i \right)^2 - \frac{K}{2N}\left( \sum_{i}s_i^2 \right)^2 \\\nonumber
&-& \sum_i (H + h_i)s_i + \Delta \sum_i s_i^2,
\end{eqnarray}
where each spin $s_i \in \{-1, 0, +1\}$. The term proportional to $J > 0$ represents a ferromagnetic interaction (set to unity hereafter), $K$ controls the bi-quadratic coupling, and $\Delta$ is a crystal field term favoring or disfavoring the $s_i = 0$ state. The $h_i$ are quenched random fields drawn independently from a Gaussian distribution with zero mean and standard deviation $R$. In the limit $K = \Delta = 0$, the model reduces to the standard RFIM.

For nonzero $K$ and $\Delta$, the presence of the $s_i=0$ state and the biquadratic coupling enrich the metastable dynamics and can lead to no-passing violation, especially in regimes with repulsive biquadratic coupling where frustration is present. The no-passing property is considered to hold for the model if, for any two spin configurations $A$ and $B$ that are 
partially ordered such that $s_i^{A} \le s_i^{B}$ for all $i$ for a given $H$, the partial ordering ($A \le B$, site-wise) is preserved during the dynamics as $H$ is increased.

\section{Glauber dynamics}
\label{Sec3}
We study the dynamics under zero-temperature single-spin-flip Glauber dynamics, where spins flip only if the energy is lowered \cite{glauber}. The external field $H$ is increased quasi-statically to generate a hysteresis loop. As $H$ increases, the system evolves through a sequence of metastable states, separated by avalanches triggered when local stability is lost. The strength of the disorder $R$ plays a key role in the structure of these avalanches. For small $R$, the system undergoes sharp collective transitions, while for large $R$ the evolution becomes smooth and gradual. At a critical value $R = R_c$, the system shows scale-invariant behavior, with avalanches following power-law distributions.

The dynamics is better understood by defining two local fields $L_1(k)$ and $L_2(k)$ associated with each spin $s_k$. They are defined as 
\begin{eqnarray}
L_1(k) &=& \frac{1}{N}\left(\sum_{i \neq k} s_i \right)+H+h_k\\
	L_2(k) &=& \frac{1}{2N} + \frac{K}{2N} \left(2\sum_{i\neq k} s^2_i + 1 \right) - \Delta
	\label{Lf2} 
\end{eqnarray}
where $N$ is the number of spins. 

The change in energy ($\delta E_k$) due to spin flip under Glauber dynamics for a randomly selected spin at site $k$, $s_k$ has 
three possibilities:

i) $s_k \rightarrow -s_k$, with $s_k =  \pm 1$,
$\delta E_k = 2 s_k L_1(k)$

ii) $s_k \rightarrow 0$, with $s_k =  \pm 1$,
$\delta E_k = s_k L_1(k) + L_2(k)$

iii) $0 \rightarrow s_k$, with $s_k =  \pm 1$, $\delta E_k = -s_k L_1(k) - L_2(k)$

\begin{figure}[t]
\centering
\includegraphics[width=0.43\textwidth]{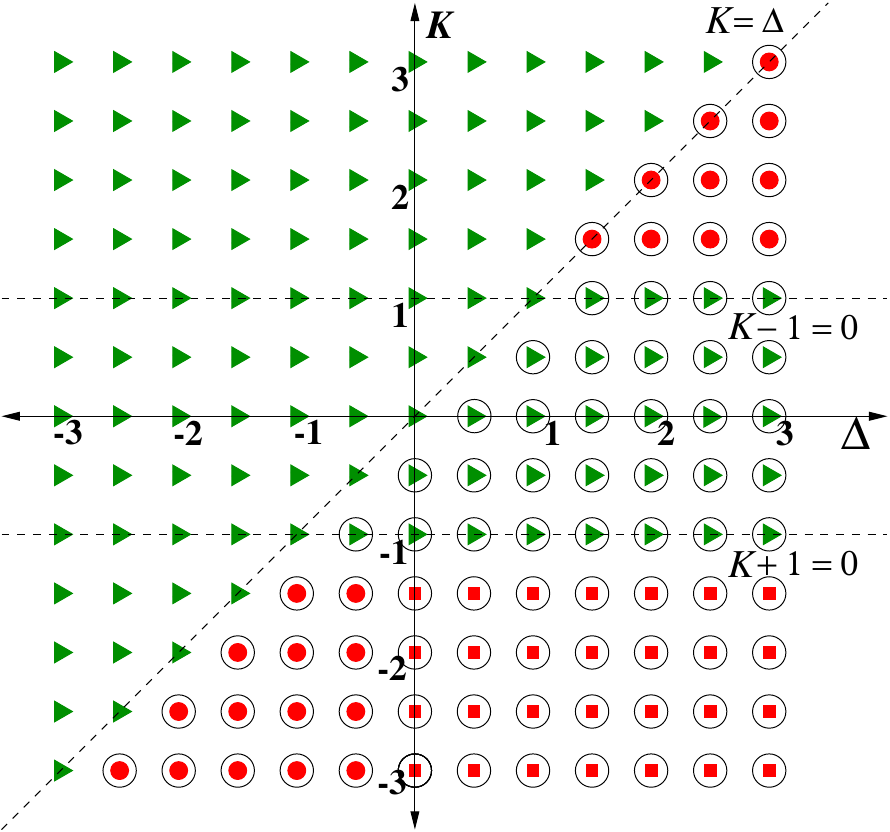}
\caption{The symbol $\textcolor{ForestGreen}{\blacktriangleright}$ is used for regions with NPP,  $\textcolor{red}{\CIRCLE}$ denotes regions with NPV with $R_c \neq 0$ and $\textcolor{red}{\blacksquare}$ denotes the region with NPV and $R_c=0$. A  
$\bigcirc$ around a symbol implies that the fraction of zero spins, $N_0 \neq 0$.The $K=\Delta$ line separates the $N_0=0$  regions from $N_0 \neq 0$. The  symbols represent the actual points in the $K-\Delta$ plane where we have explicitly verified the dynamics in simulations. }
\label{fig2}
\end{figure}

The possibility with the lowest $\delta E_k$ is chosen. We find that the spin update is independent of the present state of the spin $s_k$ (for detail see Appendix \ref{appendixA}). It depends crucially on the sign of $L_2(k)$. The magnitudes of $L_1(k)$ and $L_2(k)$ decide the actual value of the spin in an update. For,  $L_2(k)<0$, depending on the value of $L_1(k)$, the updated spin $s_k$ is given by
\begin{align}
 L_1(k) \in
	\left\{
	\begin{array}{ll}
		\left( -\infty, L_2(k) \right) & \implies s_k = -1 \\[10pt]
		\left[L_2(k), -L_2(k) \right) & \implies s_k = 0 \\[10pt]
		\left[ -L_2(k), +\infty \right) & \implies s_k = +1
	\end{array}
	\label{rule1}
	\right.
	\end{align}
Similarly for $L_2(k) \geq 0$, we get
\begin{align}
	L_1(k) \in
	\left\{
	\begin{array}{ll}
		(-\infty, 0) & \implies s_k = -1 \\[10pt]
		\left[0, \infty\right) & \implies s_k = +1
	\end{array}
	\right.
	\label{rule2}
\end{align}
Hence we find that if $L_2(k)>0$  $\forall  k$ at all times, 
then the $0$ spins play no role  and the dynamics is the same as that of the RFIM. It is easy to see that this condition always holds for $K \geq 0$ and $\Delta \leq 0$. Also, for $K<-1$ and $\Delta>K(1-1/N)$, $L_2(k)$ is negative at the start of the dynamics, since all spins are $-1$. We can hence expect $0$  spins to appear in this case. We  studied different $(K,\Delta)$  values and found that the line  $K=\Delta$  in the $(K,\Delta)$ plane separates the regions with zero and non-zero fractions of $0$ spins (see  Fig. \ref{fig2}). 

To look for regions with NPP violation, we numerically studied the $(K,\Delta)$ plane. The NPP was considered to be violated  if $1 \rightarrow 0$ or $0 \rightarrow -1$ on increasing $H$ for any $R$ at least once. We observed that NPV occurred for $\Delta>K$ provided $|K|>1$. The regions with NPP and NPV are shown in Fig. \ref{fig2}. In  the Appendix \ref{appendixB} we provide explicit examples of NPV by taking a graph with $N=3$ sites for $|K|>1$  and  $\Delta>K$. We show that NPV can lead to non abelianity for $K<-1$. For $K>1$ though the dynamics did not become non-abelian on the $N=3$ graph. We also provide numerical evidence of NPV for $N=1000$ in Appendix \ref{appendixB}. The $K<-1$  and $K>1$ with $\Delta>K$ both have NPV dynamics but they differ from each other due to the presence of competing interactions for $K<-1$ that leads to frustration.

No-passing violation implies that abelianity is not guaranteed: the final metastable state reached after an avalanche may depend on the update order. 
In the frustrated regime ($K<-1$ with $\Delta>K$), we find explicit non-abelian examples already on small graphs and confirm non-abelian behavior numerically.
For $K>1$ and $\Delta>K$, although NPP can be violated, frustration is absent and the dynamical behavior is qualitatively different.

\begin{figure*}[t]
\centering{\includegraphics[width=0.8\hsize]{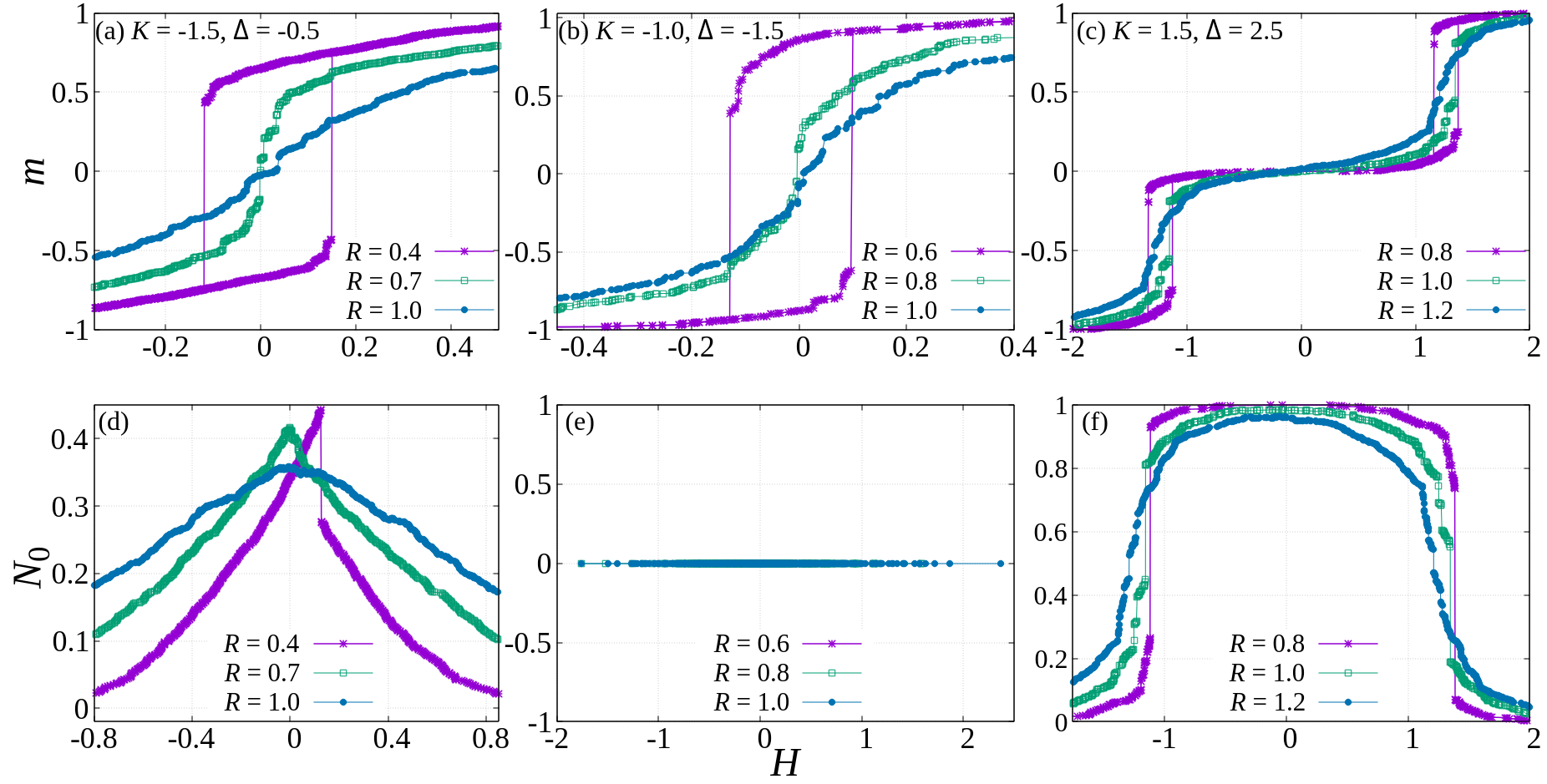}
\caption{In the first row $m-H$ is plotted for a typical realization of disorder. In each case forward plot is obtained by starting with large negative $H$ and backward plot by starting with large positive $H$. For $R<R_c$, the system shows first order hysteresis that vanishes at $R_c$. The second column shows evolution of the fraction of $0$ spins ($N_0$) in each case for  the forward plot. Specifically,  (a) and (d) are for NPV  with frustration ($K=-1.5; \Delta =-0.5$) with $R_c \approx 0.7$; (b) and (e) are for NPP ($K=-1.0; \Delta =-1.5$) with $R_c \approx 0.8$; and (c) and (f) are for NPV without frustration ($K=1.5; \Delta =2.5$) with $R_c \approx 1.0$ } \label{fig3}}
\end{figure*}

The magnetization $m$ as a function of $H$ shows first order hysteresis for $R<R_c$ and there is no hysteresis for $R>R_c$. We have plotted it for three representative values of the parameters corresponding to NPV with frustration, NPP and NPV without frustration in  Fig. \ref{fig3}. The average fraction of $0$ spins $N_0$, as a function of $H$ are also plotted. We find that the $m-H$ hysteresis plots look similar for NPP and NPV with frustration cases. For NPV without frustration,  there are two hysteresis loops that correspond to transition  from $-1$ to $0$ and $0$ to $+1$  spin state respectively. The average value of $N_0$ was found to have  different qualitative behaviour in the three regions as shown in Fig. \ref{fig3}. We also  observe  that for NPV with frustration, when $K < -1$ and $\Delta>0$ the hysteresis vanishes and there is a smooth magnetisation curve with $R_c=0$ (see  Fig.  \ref{fig2}).

\begin{figure}[t]
\centering{\includegraphics[width=1.0\hsize]{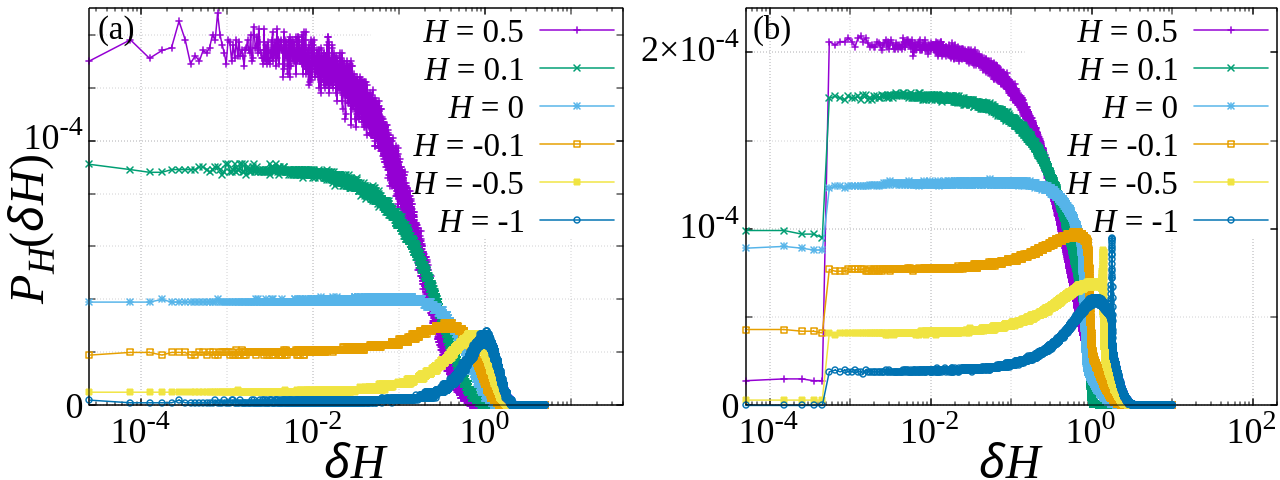}
\caption{Distribution of the field increment $\delta H$ required to destabilize the system from a metastable state at fixed $H$ at a random site, $P_H(\delta H)$ for Panel a) : NPP with $K=-1.0; \Delta =-1.5$ and Panel b) : NPV with $K=-1.5; \Delta =-0.5$ }\label{fig4}}
\end{figure}

\subsection{Distribution of  the field increment $P(\delta H)$}
To compare and study the dynamics in the NPP and NPV cases in detail, we take two sets: $(K,\Delta)=(-1.0,-1.5)$ for which the NPP holds, and $(K,\Delta)=$ $(-1.5,-0.5)$ for which the NPV occurs. The values  are chosen close enough so that there is  not much difference in their $R_c$ values (see Fig. \ref{fig3}). The $m-H$ plots also have similar behaviour in the two cases. As shown in Fig \ref{fig3}, $N_0=0$ in the case of $(K,\Delta)=(-1.0,-1.5)$. On the other hand, for $(K,\Delta)=$ $(-1.5,-0.5)$ , $N_0 \neq  0$ and shows a first order jump for $R<R_c$ and shows smooth non-monotonic behavior for $R>R_c$.

Even though the hysteresis plots look similar, we find that the $P(\delta H_{min})$ for  the two cases is strikingly different. For NPP it has a uniform distribution for small values of $\delta H_{min}$ that decays exponentially at larger values. On the other hand, for NPV the distribution shows a clear discontinuity (see Fig.  \ref{fig1}). The jump location moves left with the system size. The decay at larger values of $\delta H_{min}$ in both cases arise from regions in $m-H$ plane where $|m|$ is large and system is nearly saturated. 

We studied many different $(K,\Delta)$ values and found that $P(\delta H_{min})$ always exhibits a jump for $\Delta>K$ provided $K<-1$. At these values the dynamics is NPV and the bi-quadratic coupling competes with the ferromagnetic coupling creating frustration. The discontinuity was found to occur 
at $(|K|-1)/N$ in the distribution $P(\delta H_{min})$ as shown in Fig. \ref{fig1} (b). For illustration of the result, in the Appendix \ref{appendixC} we plot $P(\delta H_{min})$ for one more set :$(K,\Delta)=(-2.5,-1.5)$. 

For $(K,\Delta)$ values that do not satisfy the above conditions, the distribution does not have a discontinuity and is flat for small $\delta  H_{min}$ and appears similar to Fig. \ref{fig1} (a). Even for $K>1$ and $\Delta>K$ where the dynamics can violate NPP the distribution of the field increments is continuous and flat, similar to Fig. \ref{fig1} (a) (see Appendix \ref{appendixC}). Interestingly, for the system sizes explored here, the distribution is consistent with the scaling form $P(\delta H_{\min}) = N\,P(N\,\delta H_{\min})$ \cite{nampoothiri,ferre} across the investigated parameter sets.

We also looked at the distribution of the minimum increment $\delta H$ required to flip a random spin when the system is in a steady state in the presence of the external field fixed at $H$, $P_H(\delta H)$.  The discontinuity is exhibited by this distribution as well (see Fig. \ref{fig4}). We find that the value of $\delta H$  where the jump occurs is insensitive to the value of $H$. Based on these observations we unravel the dynamics of RFBEGM for different $(K,\Delta)$ below. We provide an explanation for the jump observed in the distribution for NPV by looking at the dynamics between the two  steady states. We explain the difference of the dynamics of driven RFBEGM for NPP and NPV cases in the next section.

\section{Analytical argument for the gap in the distribution}
\label{Sec4}

Let us first look at the $\delta H$ values in the pure 
model (BEGM) i.e set $h_i=0$ $\forall i$. We restrict ourselves to $\Delta>K$, as for $\Delta<K$ the dynamics is the same as that of RFIM. We start with all spins in the $-1$ state  and large negative $H$. Since there are no random fields, all the spins see the same atmosphere. The first spin to be flipped is randomly picked (we label it as $1$). The $H$ is incremented so that
its local field $L_1(k = 1)=L_2(k = 1)$.

On further infinitesimal increase in the external field $H$, the spin  $k=1$ flips to $0$ (see Eq. \ref{rule1}). Following the flip of the first spin, the local fields of the first spin remain the same, whereas the local fields of the other $N-1$ spins change their value. The local field $L_1(k \neq 1)$ of the other $N-1$ spins increases by $1/N$ and the local field $L_2(k \neq 1)$ of the other $N-1$ spins increases by $|K|/N$ for $K < 0$ and decreases by $|K|/N$ for $K \ge 0$. 

Hence, for $K  < -1$, with the first spin flipped to $0$, the local fields of other spins change so that $L_2$ is $(|K|-1)/N$ greater than $L_1$ for all the other $N-1$  spins. As a result, no other spin flips and the system reaches a steady-state  after just the flipping of the first spin. Next spin flip occurs with an  increment in $H$ of $\delta H = (|K|-1)/N$. The system again reaches the steady state after the flip of this spin. Every time a spin flips, $L_2$ for  other spins increases by an amount $|K|/N$ and the  steady state is reached after just one spin-flip. Eventually $L_2$ becomes $0$ 
and there is one large avalanche in the system in which all
the spins flip. In summary, in pure BEGM, for $K \le -1$ all avalanches are of size $1$ except the last one, which occurs when $L_2 = 0$.  

\begin{figure}[t]
\centering{\includegraphics[width=1.0\hsize]{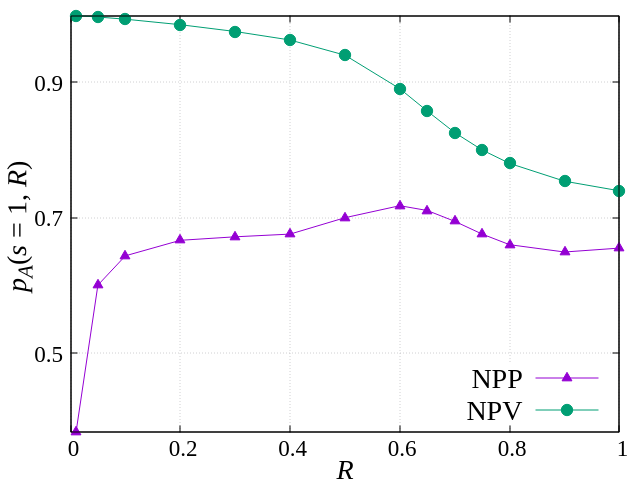}
\caption{Plot of the probability of avalanche of size $1$ (
$p_A(1,R)$) as  a function of $R$ for NPV  ($K= -1.5,\Delta= -0.5$) and NPP   ($K=-1.0,\Delta=-1.5$).}
\label{fig5}}
\end{figure}

For the RFBEGM, the local field \( L_1(k) \) is drawn from a Gaussian random field distribution with mean 0 and variance \( R \). The fields \( h_k \) are quenched at each site and remain fixed during the dynamics. The spin with the largest value of the local random field \( h_k \) is selected first. Then, the external field \( H \) is increased just enough for this spin to flip. At this point, the values of \( L_1 \) and \( L_2 \) for the remaining spins are updated in a way analogous to the BEGM.

The quenched random fields on the other sites are smaller than that of the spin that has already flipped. As a result, flipping the next spin requires an increase of at least \( \delta H_{\min} \geq (|K|-1)/N \). Consequently, for the RFBEGM, the distribution \( P(\delta H_{\min}) \) exhibits a gap (and a peak) precisely at 
\begin{equation}
\delta H_{c} = \frac{(|K|-1)}{N } 
\end{equation}
 as shown in Fig.~\ref{fig1}(b)).

Physically, this reflects the fact that in the frustrated regime a single spin flip increases the effective ``blocking'' field scale for the remaining spins by an amount of order $(|K|-1)/N$, stabilizing the system over a finite range of driving field increments.

We also examined other values of \( (K, \Delta) \) within the frustrated NPV regime of the RFBEGM, and we consistently found that a gap appears at \( \delta H_{\min} N = |K|-1 \), provided \( \Delta > K \) and \( K < -1 \). In Fig.~\ref{fig1} (b), the distribution \( P(\delta H_{\min}) \) shows the expected gap at \( (|K|-1)/N \), but also displays a small nonzero probability for events with \( \delta H_{\min} \leq (|K|-1)/N \). These correspond to the rare cases where \( L_2 \) becomes \( \geq 0 \).

For completeness, we briefly discuss two other regimes where the behavior is less rich.

First, as shown in Fig.~\ref{fig2}, when \( |K| < 1 \) or \( \Delta < K \), the system obeys No-Passing Rule. In the pure BEGM, the dynamics in this regime is reminiscent of the Ising model: a single, system-spanning avalanche occurs when \( L_1 = L_2 \). In the disordered RFBEGM, averaging over disorder realizations yields a flat distribution \( P(\delta H_{\min}) \), as in the RFIM (see Fig.~\ref{fig1}(a)).

Second, for \( K > 1 \) and \( \Delta > K \), the system enters the NPV regime, where the No-Passing Rule is violated. In this case, both bilinear and biquadratic couplings are positive and mutually reinforcing, so the system is unfrustrated. The dynamics becomes effectively decoupled between transitions from \( -1 \) to \( 0 \) and from \( 0 \) to \( +1 \), leading to two separate hysteresis loops as in the random-field Blume-Capel model (\( K = 0 \) RFBEGM)~\cite{aldrin}. Despite the violation of NPP, the distribution \( P(\delta H_{\min}) \) remains flat, as in the NPP regime (see Fig. \ref{figS5} in Appendix \ref{appendixC}).

The argument for the pure BEGM suggests that in the presence of frustration, NPV dynamics leads to many avalanches of size 1. To test this, we studied the probability of single-spin avalanches in RFBEGM for both the NPP and NPV regimes as a function of disorder strength \( R \), as shown in Fig.~\ref{fig5}. We find that for RFBEGM with NPV, probability of single spin avalanches is one at $R=0$ and even though it decreases with $R$, it consistently exhibits a much higher fraction of avalanches of size 1. In contrast, in the NPP regime, the probability of single-spin avalanches is zero at \( R = 0 \) and increases with \( R \).

\section{Avalanche size distribution}
\label{Sec5}
\begin{figure}[t]
\centering{\includegraphics[width=1.0\hsize]{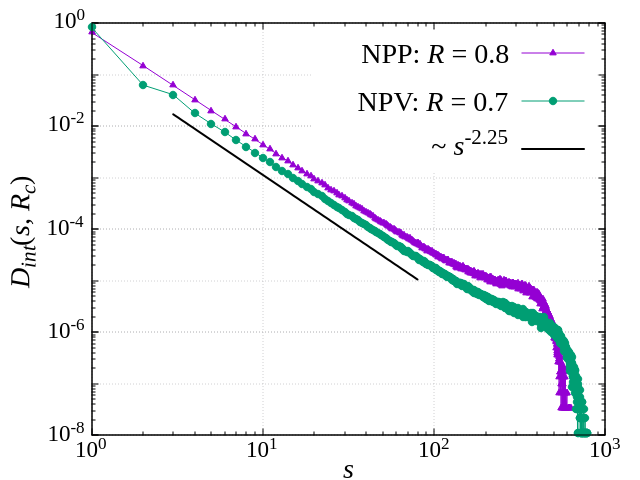}
\caption{The integrated avalanche size distributions $D_{int}(s,R_c)$ of the two cases NPP ($K=-1.0;\Delta= -1.5$) and NPV with frustration ($K=-1.5;\Delta=-0.5$) are plotted. Both distributions have a power law dependence on $s$, similar to RFIM with  $D_{int}(s,R_c) \sim s^{-2.25}$. The  data is obtained for $N=1000$ by averaging over $10^5$ realizations.}
\label{fig6}}
\end{figure}

Interestingly, this enhanced fraction of size-1 avalanches in the NPV regime does not appear to alter the exponent of the avalanche size distribution. We define $D(s,R,H)$ as the probability that the avalanche of size $s$ occurs at $H$ when the external field is increased by an infinitesimal amount and $D_{int}(s,R)$ is $D(s,H,R)$ integrated over the entire range of $H$. $D(s,R_c,H_c)$  and  $D_{int}(s,R_c)$ for RFIM varies as $s^{-1.5}$ and $s^{-2.25}$ respectively for the mean field case \cite{sethna}. In Fig.~\ref{fig6}, the integrated distribution $D_{{int}}(s, R)$  is plotted at $R_c$. For both NPP and NPV with frustration, the distribution follows the power law \( s^{-2.25} \), similar to what is observed in the RFIM. 

\begin{figure}[t]
\centering{\includegraphics[width=1.0\hsize]{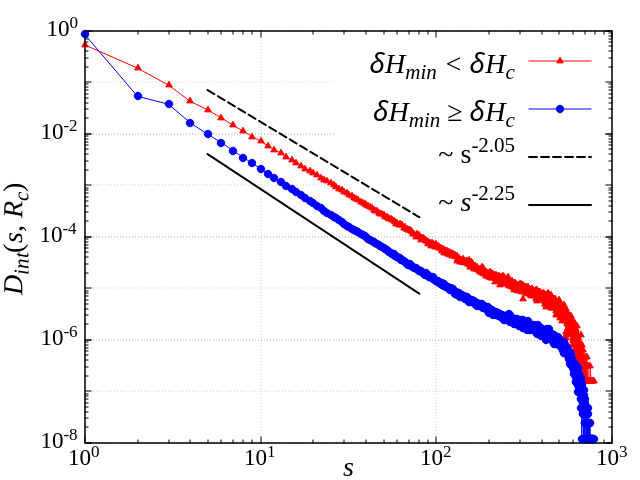}
\caption{The integrated avalanche size distributions $D_{int}(s,R_c)$ of the case NPV with frustration ($K=-1.5;\Delta= -0.5$) are plotted for $\delta H_{min} < \delta H_c$ and $\delta H_{min} \geq \delta H_c$  in red and blue respectively. The $\delta H_c= (|K|-1)/N$. The two distributions exhibit a power law but with different exponents. We find, for $\delta H_{min} \geq \delta H_c$ the exponent is $2.25$ but for $\delta H_{min} < \delta H_c$ the exponent is $2.05$. The data is obtained for $N=1000$ by averaging over $10^5$ realizations.}
\label{fig7}}
\end{figure}

We also separately looked at the  avalanche  distribution $D_{int}(s,R_c)$ of the avalanche events that occur with  $\delta H_{min}< \delta H_{c}$ and $\delta H_{min}  \ge \delta H_{c}$, where $\delta  H_c  = (|K|-1)/N$ for  the NPV. They are plotted in Fig. \ref{fig7}. We find that for $\delta H_{min} \ge\delta H_{c}$ the distribution is consistent with the exponent $2.25$, while for $\delta H_{min}< \delta H_{c}$ the effective exponent over the accessible system sizes is closer to $2.05$.We emphasize that we do not interpret this as evidence for distinct universality classes, but rather as a dynamical separation associated with the discontinuity in $\delta H_{\min}$.

\textcolor{blue}{\begin{figure*}[t]
\centering{\includegraphics[width=1.0\hsize]{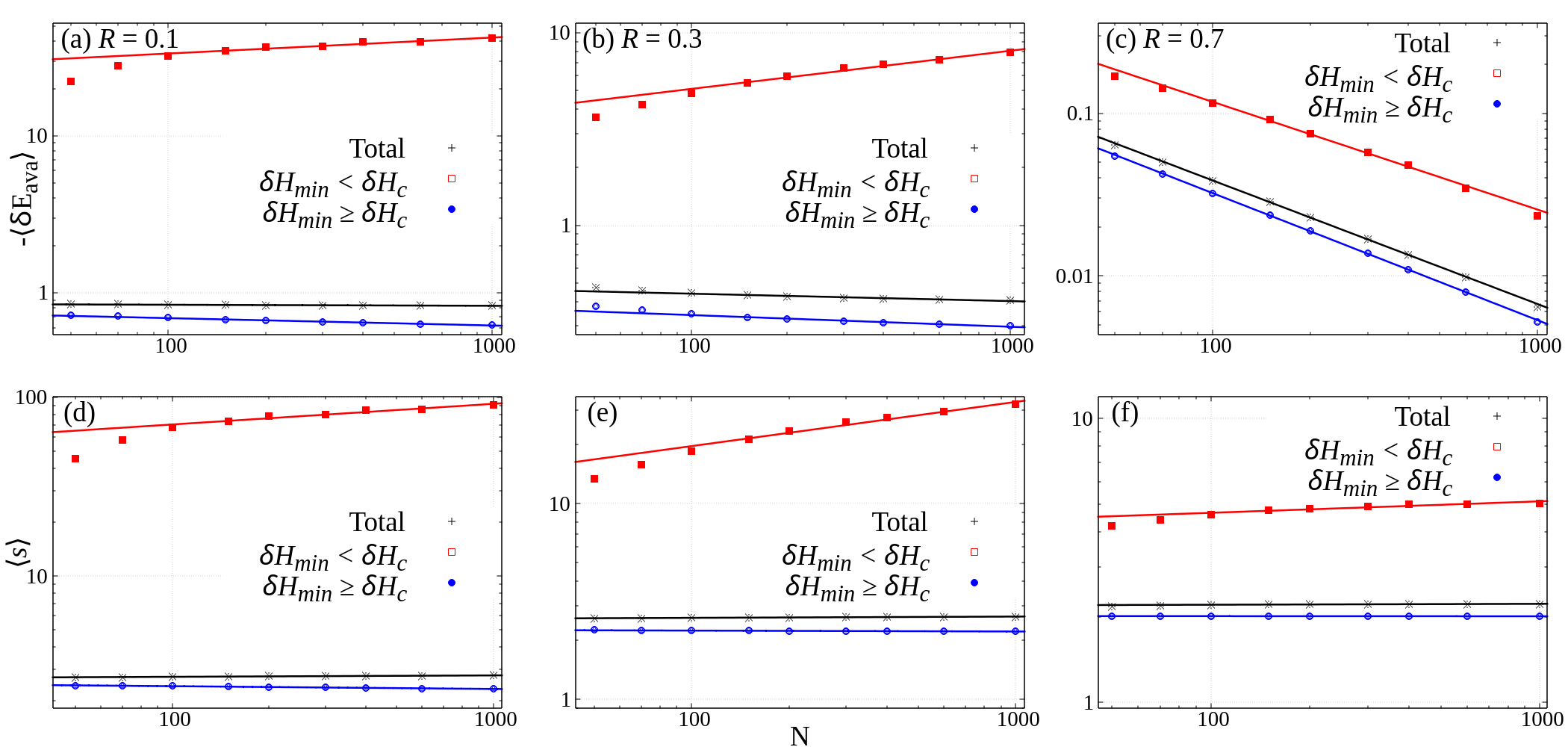}}
\caption{The plots of  $-\langle\delta E_{ava}\rangle$ and $\langle s\rangle$ over various realizations of random field distributions as a function of $N$ for events with $\delta H_{min} < \delta H_{c}$ and $\delta H_{min} \geq \delta H_{c}$ are plotted in the first and second rows in red and blue respectively for three different disorder strengths with error bars. The log-log plots of $-\langle\delta E_{ava}\rangle$ and $\langle s\rangle$ show a power-law dependence with $N$ for a given $R$. The power-law exponents for $-\langle\delta E_{ava}\rangle$ are found to be $0.10,\ 0.20,\ -0.663$ for the events with $\delta H_{min} < \delta H_{c}$ and for $\delta H_{min} \geq \delta H_{c} $ the power law exponent are roughly $-0.046,\ -0.06,\ -0.782$ for $R = 0.1, 0.3$ and $0.7$ respectively. The exponents for $\langle s\rangle$ are found to be $0.12,\ 0.23,\ 0.04$ for $\delta H_{min} < \delta H_{c}$ with $R = 0.1, 0.3, 0.7$ respectively. On the other hand for $\delta H_{min} \geq \delta H_{c} $, the $\langle s \rangle$  has  a very weak  dependence  on $N$ with the power  law exponents close to $0$. The exponents of the `Total" events shown in black have similar behaviour as that of the $\delta H_{min} \geq \delta H_{c} $ for both $-\langle\delta E_{ava}\rangle$ and $\langle s\rangle$, as there is a very small fraction of events that have $\delta H_{min} < \delta H_{c}$.}
\label{AverageDeltaE_n_NPV_R}
\end{figure*}}

\textcolor{blue}{\begin{figure}[t]
\centering{\includegraphics[width=1.0\hsize]{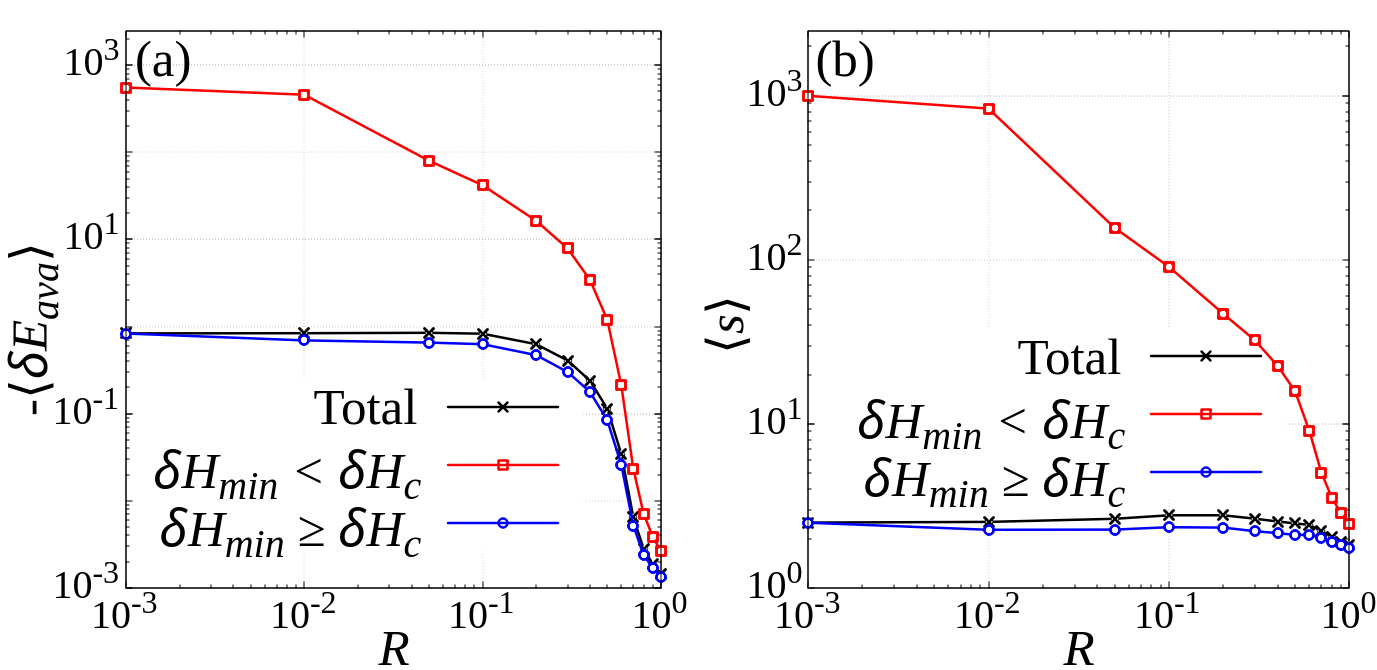}}
\caption{The average quantities  $\langle\delta E_{ava}\rangle$ and $\langle s\rangle$ as functions of $R$ for events with $\delta H_{min} < \delta H_{c}$ , $\delta H_{min} \geq \delta H_{c}$ and for all events (marked as {\it ``Total"}) are plotted for system size $N = 1000$ averaged over $5000$ realizations of the random field distributions. }
\label{Average_R_NPV}
\end{figure}}
For  $K<-1$ and $\Delta>K$ (the NPV with frustration), events with $\delta H_{min}  <  \delta H_c$  occur when $L_2<0$  and tend to  stabilize the system. This is reminiscent of phenomenology discussed in the context of localized rearrangements in amorphous solids that involve localized reconfigurations of the system \cite{karmakar, berthier,Scaling_amorphous_1}. We hence looked at the average energy released during an avalanche, $\langle \delta E_{ava} \rangle$ and average avalanche size $\langle s \rangle$ as a function of $N$ for $K=-1.5; \Delta=-0.5$. Surprisingly the quantities $\langle\delta E_{ava}\rangle$ and $\langle s\rangle$ show a power-law dependence on the system size $N$, as seen in the case of yielding of amorphous solids via plastic deformations \cite{Scaling_amorphous_1, karmakar}. 

We stress that our results are obtained in a fully connected mean-field setting, and a detailed comparison with finite-dimensional elasto-plastic models is beyond the scope of this work.

As shown in Fig. \ref{AverageDeltaE_n_NPV_R}, $-\langle  \delta E_{ava} \rangle$ for $\delta H_{min}  <  \delta H_c$ increases with $N$ for smaller $R$ and decreases with $N$ for larger $R$. Also it is larger than that of the $\delta H_{min}  \geq \delta H_c$  events for all $R$. The $\langle  s  \rangle$ also behaves similarly. The $\langle  s  \rangle$ for  the $\delta H_{min}  <  \delta H_c$  events is larger than that of the $\delta H_{min}  \geq  \delta H_c$  events for a fixed $N$ for all values of $R$ (see Fig. \ref{AverageDeltaE_n_NPV_R}). For $\delta H_{min}  \geq  \delta H_c$, $\langle s \rangle $ has a very weak dependence on $N$, indicating the localized nature of these avalanches.

The difference in the averages in the two regimes of $\delta H_{min}$ decreases as $R$ increases as shown in Fig.\ref{Average_R_NPV}. This suggests that the disorder tends to balance the effect of frustration. We have also plotted  the  average values when the average is done over all  values of $\delta H_{min}$ in black. These averages behave  similar to $\delta H_{min}  \geq  \delta H_c$   events for all  $R$ and $N$  studied. This is expected, as the $\delta H_{min}  <  \delta H_c$  events have a very low probability  of occurence.

\section{Discussion}
\label{Sec6}

The BEGM is known for its rich behavior, as it includes both bilinear and biquadratic couplings, as well as a crystal field term - unlike the Ising model, which involves only bilinear interactions. In this work we studied the zero-temperature quasi-static dynamics of the RFBEGM across its parameter space in the fully connected (mean-field) limit. 
We showed that no-passing violation alone does not necessarily produce a qualitative change in the field-increment statistics. 
In contrast, when no-passing violation is combined with frustration induced by a repulsive biquadratic coupling, the distribution of minimal field increments develops a robust discontinuity, whose location can be predicted analytically and is confirmed numerically. 
We interpret this discontinuity as a dynamical diagnostic of frustration-induced blocking in non-abelian avalanche dynamics, and an interesting direction for future work is to explore how these signatures evolve in finite dimensions.

The excitation stress distribution in plastic crystals \cite{ovaska}, the plastic strain distribution in amorphous materials \cite{lin}, and the distribution of field increments for spin glass models \cite{skmodel2,skmodel1} all show pseudo-gap and a peak at a non-zero value of the increment under quasi-static driving. Pseudo-gap makes sure that a plastic event doesn't result in significant rearrangements. The $P(\delta H_{min})$ distribution for 
RFBEGM in the frustrated regime  with non abelian dynamics also shows a threshold effect and peaks at a finite value of $\delta H_{min}$.

The discontinuity in the gap distribution separates the avalanche events into two different kinds. The events with $\delta H_{min} \ge \delta H_c$ are more localized with weaker dependence on $N$ and are related to frustration induced blocking. A detailed study of the criticality in the two regimes especially on finite dimensional lattices would be useful. It would also be interesting to look at the dynamics at finite temperature \cite{yao} to see if the temperature acts similar to $R$ and reduces the impact of repulsive interaction.

Both NPP and NPV regimes leave observable signatures in the average value of \( N_0 \), even though they are indistinguishable from one another in the hysteresis loops. It would be particularly intriguing to explore whether some kind of frustration such as the one introduced by $K<-1$ and $\Delta>K$ could be used to construct a model of elastic manifolds in random media that exhibits properties different from those of the standard depinning transition.

{\it Acknowledgement : }
AR and S acknowledge the International Centre for Theoretical Sciences (ICTS) and CEFIPRA for the workshop - Indo-French Workshop on ``Classical and quantum dynamics in and out of equilibrium systems" (code: ICTS/IFWCQM2024/12), for  providing platform for discusssion. S acknowledges Smarajit Karmakar for discussions. S would like to acknowledge support from the ICTP through the Associates Programme and from the Simons Foundation through grant number 284558FY19.

\appendix

\label{appendix}

\section{Derivation of spin update rules}
\label{appendixA}
The change in energy of the system when $s_k\in \{-1,+1\}$ flips to $s^\prime_k\in \{-1,+1\}$ is
\begin{align}\delta E_{s_k\rightarrow s^\prime_k}&= 2s_k \bigg[\frac{J}{N} \sum_{i \neq k} s_i + H + h_k \bigg]\\
		&= 2s_kL_1(k)
	\label{Change_E_stos}
\end{align}

Similarly, the energy when $s_k \in \{-1,1\}$ flips to $0$ is 

\begin{eqnarray}
\delta E_{s_k \rightarrow 0} &=& s_k \bigg[\frac{J}{N} \sum_{i \neq k} s_i + H + h_k \bigg]+ \\\nonumber
		&& \bigg[\frac{J}{2N} + \frac{K}{2N} \bigg(2 \sum_{i \neq k} s_i^2 + 1 \bigg) - \Delta \bigg]\\
		&=& s_kL_1(k) + L_2(k)
	\label{Change_E_sto0}
\end{eqnarray}

Energy when $0$ flips to $s_k\in \{-1,1\}$ is
\begin{eqnarray}
		\delta E_{0 \rightarrow s_k} &=& -s_k \bigg[\frac{J}{N} \sum_{i \neq k} s_i + H + h_k \bigg]-\\\nonumber 
		&& \bigg[\frac{J}{2N} + \frac{K}{2N} \left( 2 \sum_{i \neq k} s_i^2 + 1 \right) - \Delta \bigg]\\
		&=& -s_kL_1(k) - L_2(k)
	\label{Change_E_0tos}
\end{eqnarray}
The spin flips to the value which most decreases the energy of the system.\\
Here,  $L_1(k)$ and $L_2(k)$ are local fields associated with $k^{th}$ spin. They are defined as: 
\begin{equation}
	L_1(k) = \frac{J}{N}\left(\sum_{i\neq k} s_i \right) + H + h_k
	\label{Lf1} 
\end{equation}

\begin{equation}
	L_2(k) = \frac{J}{2N} + \frac{K}{2N} \left(2\sum_{i\neq k} s^2_i + 1 \right) - \Delta
	\label{Lf2} 
\end{equation}
The flipping of a spin crucially depends on the sign of the local field $L_2$.

\subsection{$L_2(k) < 0$} 
For $s_k = -1$, $\delta E_{-1 \rightarrow s_k^\prime } $ is
$$\delta E_{-1 \rightarrow -1 } = 0$$ $$\delta E_{-1 \rightarrow 0 } = -L_1(k)+L_2(k)$$ $$\delta E_{-1 \rightarrow +1 } = -2L_1(k)$$
The value of $s_k^\prime$ that results in lowest energy depends only on the magnitude of $L_1(k)$ and is given by
\begin{align}
	L_1(k) \in
	\left\{
	\begin{array}{ll}
		\left( -\infty, L_2(k) \right) & \implies s^\prime_k = -1 \\[10pt]
		\left[ L_2(k), -L_2(k) \right) & \implies s^\prime_k = 0 \\[10pt]
		\left[ -L_2(k), +\infty \right) & \implies s^\prime_k = +1
	\end{array}
	\right.
	\label{eq_case1:-1}
\end{align}
Similarly, for $s_k = 0$, $\delta E_{0 \rightarrow s_k^\prime } $ is
$$\delta E_{0 \rightarrow -1 } = L_1(k)-L_2(k)$$
$$\delta E_{0 \rightarrow 0 } = 0 $$
$$\delta E_{0 \rightarrow +1 } = -L_1(k)-L_2(k)$$
Thus,
\begin{align}
	L_1(k) \in
	\left\{
	\begin{array}{ll}
		\left( -\infty, L_2(k) \right) & \implies s^\prime_k = -1 \\[10pt]
		\left[ L_2(k), -L_2(k) \right) & \implies s^\prime_k = 0 \\[10pt]
		\left[ -L_2(k), +\infty \right) & \implies s^\prime_k = +1
	\end{array}
	\right.
	\label{eq_case1:0}
\end{align}

For $s_k = +1$, $\delta E_{+1 \rightarrow s_k^\prime } $ is

$$\delta E_{+1 \rightarrow -1 } = 2L_1(k)$$
$$\delta E_{+1 \rightarrow 0 } = L_1(k)+L_2(k) $$
$$\delta E_{+1 \rightarrow +1 } = 0$$
Thus,
\begin{align}
	 L_1(k) \in
	\left\{
	\begin{array}{ll}
		\left( -\infty, L_2(k) \right) & \implies s^\prime_k = -1 \\[10pt]
		\left[ L_2(k), -L_2(k) \right) & \implies s^\prime_k = 0 \\[10pt]
		\left[ -L_2(k), +\infty \right) & \implies s^\prime_k = +1
	\end{array}
	\right.
	\label{eq_case1:+1}
\end{align}

\subsection{$L_2(k) \geq 0$} 
For $s_k = -1$, $\delta E_{-1 \rightarrow s_k^\prime } $ is

$$\delta E_{-1 \rightarrow -1 } = 0$$
$$\delta E_{-1 \rightarrow 0 } = -L_1(k)+L_2(k) $$
$$\delta E_{-1 \rightarrow +1 } = -2L_1(k)$$
Thus,
\begin{align}
	L_1(k) \in
	\left\{
	\begin{array}{ll}
		(-\infty, 0) & \implies s^\prime_k = -1 \\[10pt]
		\left[0, \infty\right) & \implies s^\prime_k = +1
	\end{array}
	\right.
	\label{eq_case2:-1}
\end{align}

For $s_k = 0$, $\delta E_{0 \rightarrow s_k^\prime } $ is 

$$\delta E_{0 \rightarrow -1 } = L_1(k)-L_2(k)$$
$$\delta E_{0 \rightarrow 0 } = 0 $$
$$\delta E_{0 \rightarrow +1 } = -L_1(k)-L_2(k)$$
Thus,
\begin{align}
	L_1(k) \in
	\left\{
	\begin{array}{ll}
		(-\infty, 0) & \implies s^\prime_k = -1 \\[10pt]
		\left[0, \infty\right) & \implies s^\prime_k = +1
	\end{array}
	\right.
	\label{eq_case2:0}
\end{align}

For $s_k = +1$, $\delta E_{-1 \rightarrow s_k^\prime } $ is 
$$\delta E_{+1 \rightarrow -1 } = 2L_1(k)$$
$$\delta E_{+1 \rightarrow 0 } = L_1(k)+L_2(k) $$
$$\delta E_{+1 \rightarrow +1 } = 0$$

Thus,
\begin{align}
 L_1(k) \in
	\left\{
	\begin{array}{ll}
		(-\infty, 0) & \implies s^\prime_k = -1 \\[10pt]
		\left[0, \infty\right) & \implies s^\prime_k = +1
	\end{array}
	\right.
	\label{eq_case2:+1}
\end{align}

\section{No passing violation and non-abelianity of the dynamics}
\label{appendixB}
\begin{figure}[ht]
		\centering
		\includegraphics[width=0.3\textwidth]{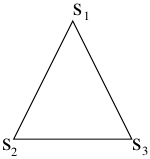}
		\caption{\centering A 3-site fully connected graph}
		\label{fig:triangle_graph}
	\end{figure}

	The no-passing property (NPP) is violated for $\Delta > K$ $\forall |K| > 1$. In this section, we show no passing violation (NPV) and non-abelianity via an example for a 3-site fully connected graph (Fig. \ref{fig:triangle_graph}). We also present numerical evidence of NPP violating spin flips for larger $N$. We can fix one of the couplings, and hence fix $J=1$.
	
	 \subsection{Case 1:  $K<-1$ }
	Assume that the system is in a steady state with $\{s_1,s_2,s_3\} = \{-1,0,0\}$. The local fields associated with different sites $s_1, s_2, s_3$ are
\begin{eqnarray}
	L_1(1) &=& H + h_1;~~~ L_2(1) = \frac{1}{6} + \frac{K}{6} - \Delta\\
	L_1(2) &=& -\frac{1}{3} + H + h_2;~~~ L_2(2) = \frac{1}{6} + \frac{K}{2} - \Delta\\
	L_1(3) &=& -\frac{1}{3} + H + h_3;~~~ L_2(3) = \frac{1}{6} + \frac{K}{2} - \Delta
	\end{eqnarray}	
Let us pick $s_1$ and try to update it. If $K$ and $\Delta$ are such that $L_2(1)$ is positive, then on increasing $H$, $s_1$ will flip to $1$ when $L_1(1)$ become greater than $0$.  System  at this value of $H$  either will already be in a steady state  or in case $s_2$ or $s_3$  or both also become unstable, then since $L_2(2)$  and $L_3(2)$ are also positive the spins will slip respecting the NPP.

But in the case where the values of $K$ and $\Delta$ are such that $L_2(1)<0$, then on increasing $H$, the spin $s_1$ will flip to $0$ when $H$ is such that $L_2(1)  <  L_1(1) < -L_2(1)$. This will make the local fields for $s_2$, $L_1(2)$ to increase by $1/N$ and $L_2(2)$ to increase by $|K|/N$. Similarly, the local fields for $s_3$, $L_1(3)$ increases by $1/N$ and $L_2(3)$ increases by $|K|/N$. Now since $|K|>1$, there is a possibility that the $h_2$ and $h_3$ are such that, $L_1(2)<L_2(2)$ and $L_1(3)<L_2(3)$  at this stage  and hence both $s_2$ and $s_3$  are unstable. 

If $s_2$ is picked for update, then $s_2$ will flip to $-1$, causing a NPP violation (NPV). After $s_2$ flips, the spin configuration will  be $\{0,-1,0\}$. This will  now  stabilize $s_3$ and it will continue to stay $0$. The final configuration reached in this case is then $\{s_1,s_2,s_3\} = \{0,-1,0\}$.

On the other hand if $s_3$ is picked for update first, the final steady state configuration reached will  be $\{s_1,s_2,s_3\} = \{0,0,-1\}$. Since the steady state depends on the order of update, the dynamics is non-abelian.
    
\subsection{Case 2:   $K>1 $}

	Assume that the system is in a steady state with $\{s_1,s_2,s_3\} = \{-1,1,1\}$. The local fields associated with different sites $s_1, s_2, s_3$ are 
	\begin{eqnarray}
	L_1(1) &=& \frac{2}{3} + H + h_1;~~~ L_2(1) = \frac{1}{6} + \frac{5K}{6} - \Delta\\
	L_1(2) &=& H + h_2;~~~ L_2(2) = \frac{1}{6} + \frac{5K}{6} - \Delta\\
	L_1(3) &=& H + h_3;~~~ L_2(3) = \frac{1}{6} + \frac{5K}{6} - \Delta
\end{eqnarray}
We first increase $H$ so that $L_1(1)>L_2(1)$, making $s_1$ unstable.  This will make $s_1$ flip to $0$. As a result the local fields for $s_2$ and $s_3$, $L_1(2)$ and $L_1(3)$ increase by $1/N$, and $L_2(2)$ and  $L_2(3)$ decrease by $K/N$. Let us assume that $h_2$ is such that $s_2$ becomes unstable with $L_1(2) < -L_2(2)$.  This will make $s_2$ flip to $0$, making the local fields for $s_1$ and $s_3$, $L_1(1)$ and $L_1(3)$ decrease by $1/N$, and $L_2(1)$ and $L_2(3)$ decrease by $K/N$. Now, let us assume that $h_3$ is large enough such that $s_3$ remains stable being $+1$, with $L_1(3) > -L_2(3)$, even after the relaxation of $s_1$ and $s_2$. Hence, a steady state is reached after the flipping of $s_1$ and $s_2$. This new steady state is $\{0,0,1\}$, which violates the NPP.

 

\begin{figure}[t]
\centering
\includegraphics[width=0.9\hsize]{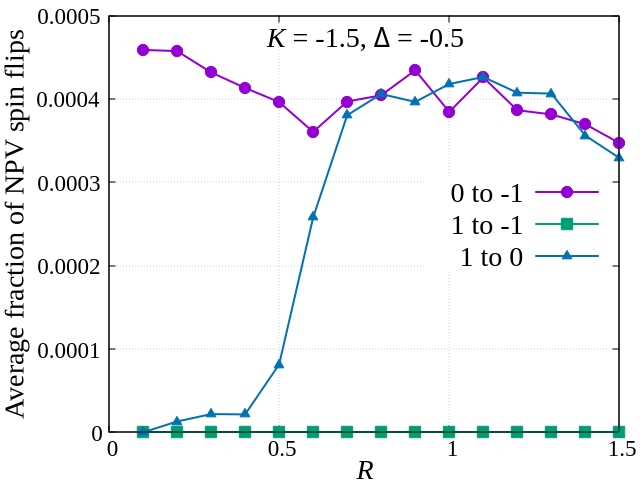}
	\caption{Fraction of NPV spin flips in the presence of frustration. Data is obtained by averaging over $1000$ disorder realizations for $N=1000$.}
    \label{figS2}
\end{figure}

\begin{figure}[t]
\centering
\includegraphics[width=0.9\hsize]{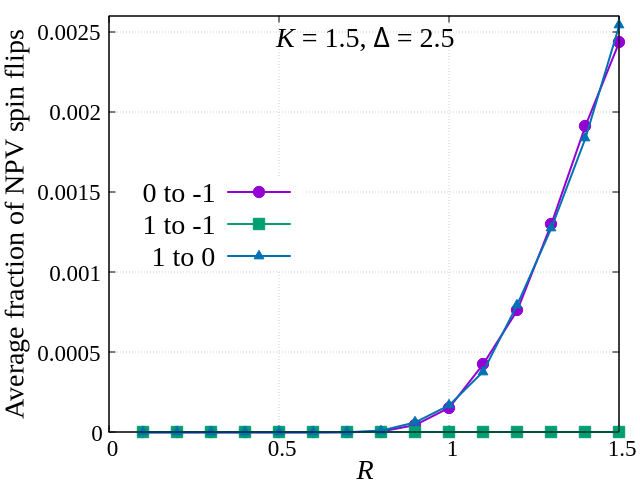}
	\caption{Fraction of NPV spin flips for  $K>1$. Data is obtained by averaging over $1000$ disorder realizations for $N=1000$.}\label{figS3}
\end{figure}

\begin{figure}[t]
\centering{\includegraphics[width=0.9\hsize]{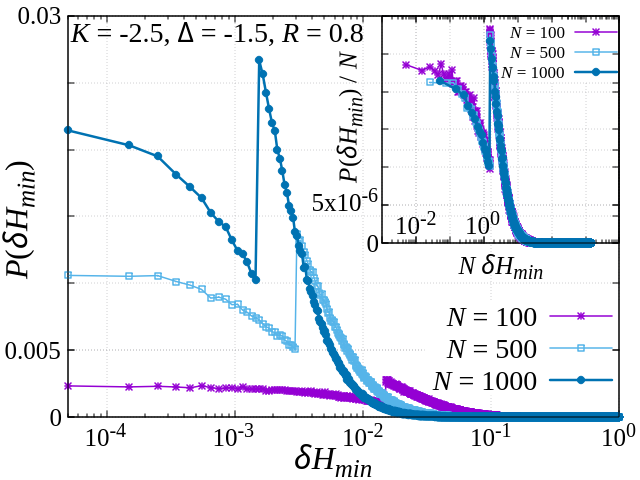}
	\caption{Distribution of the minimal field increment $\delta H_{\min}$ required to destabilize the system from a metastable state for $(K,\Delta)=$ $(-2.5, -1.5)$. The discontinuity occurs at $(|K|-1)/N$.}\label{figS4}}
\end{figure}

\begin{figure}[t]
\centering{\includegraphics[width=0.9\hsize]{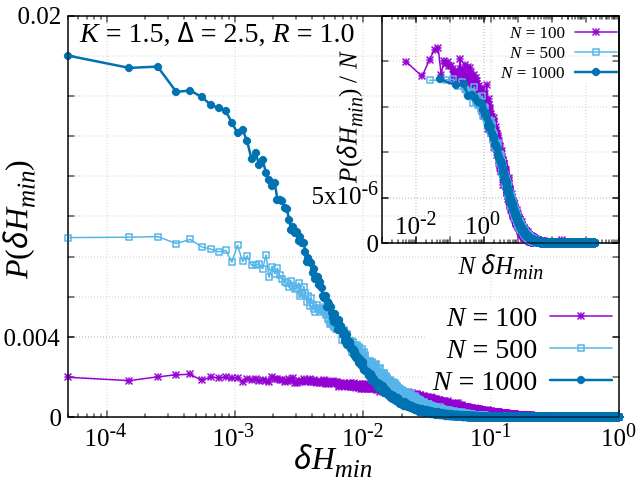}}
	\caption{Distribution of the minimal field increment $\delta H_{\min}$ required to destabilize the system from a metastable state for $(K,\Delta)=$ $(1.5, 2.5)$.} \label{figS5}
\end{figure}


\subsection{NPV for larger $N$ via simulations}
In the simulations, we collected the statistics of the fraction of spin flips which violates the NPP. These are plotted, respectively, in Fig. \ref{figS2} and Fig. \ref{figS3} for $K=-1.5,\Delta=-0.5$  and $K=1.5,\Delta=2.5$ for $N=1000$. One finds that while NPV violating spin flips are seen at all values of $R$ in the frustrated case with $K<-1$, for $K>1$  and $\Delta>K$, it occurred at higher values of R. This is  because below  $R_c$, the dynamics from $-1$ to $0$ and $0$ to $1$ is largely independent.

\section{Distribution of field increments between successive avalanches}
\label{appendixC}
The distribution of the minimum field increment $\delta H_{min}$ required to trigger the next instability shows the clear signature of NPV  with frustration in the form of a discontinuity. We give analytical arguments for the location of the discontinuity, which give the location of the discontinuity to be $\delta H_{min} N=|K|-1$, valid for $K<-1$  
with $\Delta >K$. In Fig.  \ref{figS4},  we illustrate it with one more example.

For  $K>1$ and $\Delta>K$ even though the NPP is violated, there is no frustration. The distribution continues to be flat, without a discontinuity,  as shown via an example in Fig. \ref{figS5}.

\end{document}